\documentstyle{mn}
\newif\ifAMStwofonts

\ifoldfss
  \ifCUPmtlplainloaded \else
    \NewTextAlphabet{textbfit} {cmbxti10} {}
    \NewTextAlphabet{textbfss} {cmssbx10} {}
    \NewMathAlphabet{mathbfit} {cmbxti10} {} 
    \NewMathAlphabet{mathbfss} {cmssbx10} {} 
  \fi
  \ifAMStwofonts
    \ifCUPmtlplainloaded \else
      \NewSymbolFont{upmath} {eurm10}
      \NewSymbolFont{AMSa} {msam10}
      \NewMathSymbol{\upi}     {0}{upmath}{19}
      \NewMathSymbol{\umu}     {0}{upmath}{16}
      \NewMathSymbol{\upartial}{0}{upmath}{40}
      \NewMathSymbol{\leqslant}{3}{AMSa}{36}
      \NewMathSymbol{\geqslant}{3}{AMSa}{3E}

    \fi
  \fi
\fi 

\ifnfssone
  \newmathalphabet{\mathit}
  \addtoversion{normal}{\mathit}{cmr}{m}{it}
  \addtoversion{bold}{\mathit}{cmr}{bx}{it}
  \newmathalphabet{\mathbfit} 
  \addtoversion{normal}{\mathbfit}{cmr}{bx}{it}
  \addtoversion{bold}{\mathbfit}{cmr}{bx}{it}
  \newmathalphabet{\mathbfss} 
  \addtoversion{normal}{\mathbfss}{cmss}{bx}{n}
  \addtoversion{bold}{\mathbfss}{cmss}{bx}{n}
  \ifAMStwofonts
    \ifCUPmtlplainloaded \else
      \UseAMStwoboldmath
      \makeatletter
      \new@mathgroup\upmath@group
      \define@mathgroup\mv@normal\upmath@group{eur}{m}{n}
      \define@mathgroup\mv@bold\upmath@group{eur}{b}{n}
      \edef\UPM{\hexnumber\upmath@group}
      \new@mathgroup\amsa@group
      \define@mathgroup\mv@normal\amsa@group{msa}{m}{n}
      \define@mathgroup\mv@bold\amsa@group{msa}{m}{n}
      \edef\AMSa{\hexnumber\amsa@group}
      \makeatother
      \mathchardef\upi="0\UPM19
      \mathchardef\umu="0\UPM16
      \mathchardef\upartial="0\UPM40
      \mathchardef\leqslant="3\AMSa36
      \mathchardef\geqslant="3\AMSa3E
    \fi
  \fi
\fi 

\ifnfsstwo
  \DeclareMathAlphabet{\mathbfit}{OT1}{cmr}{bx}{it}
  \SetMathAlphabet\mathbfit{bold}{OT1}{cmr}{bx}{it}
  \DeclareMathAlphabet{\mathbfss}{OT1}{cmss}{bx}{n}
  \SetMathAlphabet\mathbfss{bold}{OT1}{cmss}{bx}{n}
  \ifAMStwofonts
    \ifCUPmtlplainloaded \else
      \DeclareSymbolFont{UPM}{U}{eur}{m}{n}
      \SetSymbolFont{UPM}{bold}{U}{eur}{b}{n}
      \DeclareSymbolFont{AMSa}{U}{msa}{m}{n}
      \DeclareMathSymbol{\upi}{0}{UPM}{"19}
      \DeclareMathSymbol{\umu}{0}{UPM}{"16}
      \DeclareMathSymbol{\upartial}{0}{UPM}{"40}
      \DeclareMathSymbol{\leqslant}{3}{AMSa}{"36}
      \DeclareMathSymbol{\geqslant}{3}{AMSa}{"3E}
    \fi
  \fi
\fi 

\ifCUPmtlplainloaded \else
  \ifAMStwofonts \else 
    \def\upi{\pi}
    \def\umu{\mu}
    \def\upartial{\partial}
  \fi
\fi

\title[Star Clusters
in the Large Magellanic Cloud]
{Hierarchical Star Formation from 
the Time-Space Distribution of Star Clusters
in the Large Magellanic Cloud}

\author[Y. Efremov and B. Elmegreen]
  {Yuri~N.~Efremov$^1$ and Bruce~G.~Elmegreen$^2$\\
  $^1$ MSU, P.K.Sternberg Astronomical Institute, Moscow 119899\\
  $^2$ IBM Research Division, T.J. Watson Research Center,
        P.O. Box 218, Yorktown Heights, NY 10598}

\date{Accepted  13 May 1998.
      Received 25 March 1998;
      in original form 18 December 1997}

\pagerange{\pageref{firstpage}--\pageref{lastpage}}
\pubyear{1998}

\begin{document}

\maketitle

\label{firstpage}

\begin{abstract}
The average age difference between pairs of star clusters in the Large
Magellanic Clouds increases with their separation as the $\sim0.35$ power.
This suggests that star formation is hierarchical in
space and time.  Small regions form stars quickly and large regions,
which often contain the small regions, form stars over a longer
period.   A similar result found previously for Cepheid variables
is statistically less certain than the cluster result. 
\end{abstract}


\section{Introduction}

A previous study of the positions and ages of Cepheid variables
in the Large Magellanic Cloud (Elmegreen \& Efremov 1996) found
that Cepheids closer to each other were also more likely to have
the same age, as if star formation proceeds in small regions
faster than in large regions.  Because Cepheids are relatively
old ($>27$ Myr in the LMC sample, with most around 100 Myr)
and could have drifted slightly
from their point of origin, and because Cepheid ages are
rather uncertain, we repeat this LMC study here using star clusters.
The Bica et al. (1996) catalog of clusters in the LMC
contains 590 members with accurate positions and UBV colors, and these colors
have been converted to age by Girardi et al. (1995). 
Thus we can use the Bica et al. clusters to study the
hierarchical properties of star formation in the LMC. 

\section{Cluster age difference versus separation}

We consider all pairs of clusters in the LMC for clusters
within certain age ranges, and determine the average age
difference $\Delta t$ among these clusters as a function of their 
separation $S$ for regular
intervals of separation. 
Separation is defined to be the
deprojected distance between the two clusters, considering zero depth
to the LMC and an inclination of 33$^\circ$ (Luk \& Rohlfs 1992).  
For the age difference,
the absolute value is used to avoid negative numbers. 

\begin{figure}
\vspace{3.2in}
\includegraphics{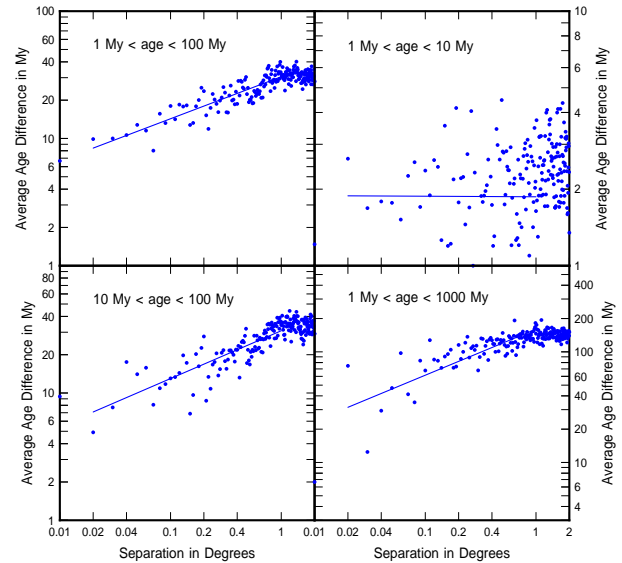}
\caption{The average age differences between pairs of star clusters are plotted
versus their deprojected angular separations for clusters in the LMC. Four intervals
of cluster ages are considered. For the larger age intervals, the clusters show a correlation
in the sense that close clusters have more similar ages than distant clusters.}
\label{fig:clustersreal}
\end{figure}

The results are shown in Figure \ref{fig:clustersreal}.
Four age intervals are considered: 1 to 100 Myr, 
1 to 10 Myr, 10 to 100 Myr, and 1 to 1000 Myr. 
The average age difference between pairs of 
clusters increases systematically
with their spatial separation. 
The number of clusters within these age intervals
is 337, 93, 244, and 526, respectively.   

\begin{figure}
\vspace{3.2in}
\includegraphics{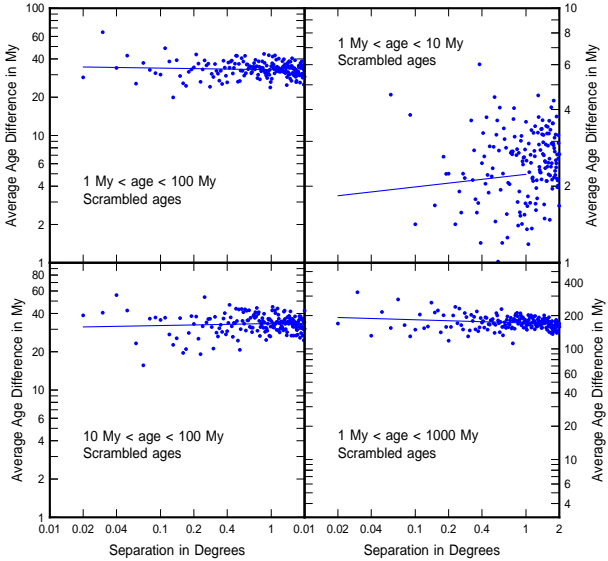}
\caption{The same as in figure 1 but with cluster ages scrambled
randomly among all of the clusters in the sample.
The correlation disappears for randomized ages, suggesting that
the trend found for the real data is statistically significant.}  
\label{fig:clusrandom} 
\end{figure}

Figure \ref{fig:clusrandom} 
shows a similar study using the same 590 cluster positions, 
but now with ages that are mixed up randomly in order to check
purely statistical effects
(these are not purely random ages, but the real cluster ages
that are reassigned randomly to 
different clusters).  The correlations disappear
for this random sample, i.e., the average age difference is 
the same regardless of separation.

\begin{figure}
\vspace{3.2in}
\includegraphics{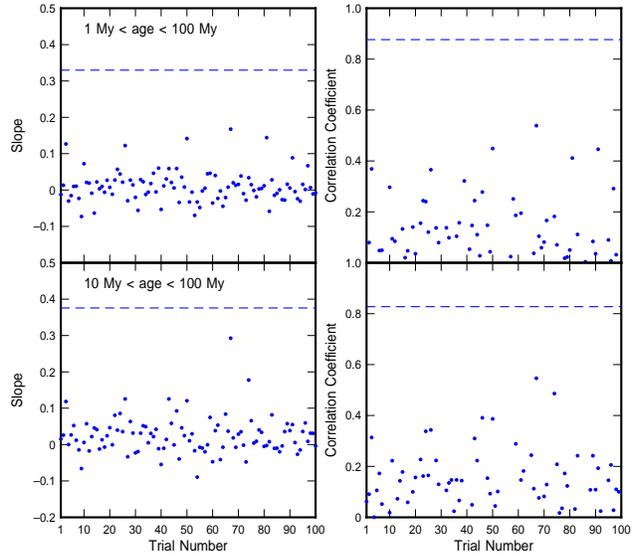}
\caption{{\it Left:} The slope for the least squares fit to the regression of 
average age difference versus cluster separation is shown as a point for each of 
100 trials
with randomly scrambled age data and real cluster positions.  
The slope has an average value of zero because
there is no $\Delta t-S$ correlation for the randomized data.
The slope for the real data, indicated by a dashed line, 
is clearly larger than the rms slopes for the random data. 
{\it Right:} The correlation coefficients for the regressions made from randomized
data are shown as points, and the correlation coefficient for the real data
is a dashed line. The correlation coefficient for the real data is large, indicating 
that the cluster $\Delta t-S$ correlation is statistically significant. 
Two intervals of cluster age are considered.} 
\label{fig:randomtrials}
\end{figure}

Figure \ref{fig:randomtrials} gives the results of 100 random trials 
using the real cluster positions and randomly mixed ages
for ages in the ranges 1-100 Myr (top) and 10-100 Myr (bottom).
Each point on the left gives the least-squares fit to the slope of the 
$\Delta t-S$ correlation (evaluated in the separation 
interval from 0.01 to 1 degree), 
and each point on the
right gives the correlation coefficient
measuring the goodness of fit (1=perfect fit).  The dashed lines
represent the slopes and correlation coefficients for the
real data, where the clusters have their real ages, as shown in Figure 1. 

Figure \ref{fig:randomtrials} indicates that the slopes of the $\Delta t-S$
correlations for the real data in the 1-100 Myr and 10-100 Myr age intervals
differ from random noise in a statistically
significant way. The slope is much higher than the slopes in the
random samples, and the correlation coefficient
for the real data is good. 

The least-squares fits in the $S=0.01-1$ degree separation
range for the significant $\Delta t-S$ relations
shown in figure \ref{fig:clustersreal} are:
\begin{equation}
\log \Delta t_(yr) = 7.48+0.33 \log S(deg.),\;\;\;(1-100\;{\rm Myr})
\end{equation}
\begin{equation}
\log \Delta t_(yr) = 7.49+0.38 \log S(deg.),\;\;\;(10-100\;{\rm Myr})
\end{equation}
\begin{equation}
\log \Delta t_(yr) = 8.20+0.42 \log S(deg.),\;\;\;(1-1000\;{\rm Myr})
\end{equation}
The correlation coefficients for these three fits are
0.88, 0.83, and 0.82, respectively; the numbers of cluster pairs
are 5509, 2849, and 10614. 
The 1-10 Myr data has too few clusters (470 pairs) to be statistically significant
(correlation coefficient: $-0.0046$). 

We conclude from these figures that clusters in the LMC 
form in a hierarchical sequence in which the duration of
star formation in a region scales with the $\sim 0.35$ power of the 
region size over scales ranging from at least 15 pc (0.02$^\circ$) to 
780 pc (1$^\circ$) at a distance of 45 kpc 
(Berdnikov, Vozyakova, \& Dambis 1996;
Efremov 1997; Efremov, Schilbach, \& Zinnecker 1997; Fernley et al. 1998).
For the 1-100 Myr age range in this $0.01-1$ degree
separation range,  this correlation is
\begin{equation}
\Delta t({\rm Myr}) \sim 3.3S({\rm pc})^{0.33}.
\label{eq:tspc}
\end{equation}

\section{Age difference versus separation for Cepheid variables}

A similar analysis is now applied to Cepheids in the LMC, as in 
Elmegreen \& Efremov (1996), but now we use an age calibration
that is consistent with the ages of the clusters. 
The ages of Cepheids come from their periods, because both the period
and the age of a Cepheid correlate with stellar mass.
The original period-age relation came from Efremov (1978)
and did not consider overshooting in the stellar evolution models. 
The ages of the clusters used in the previous section do consider 
overshooting,
so we have to recalibrate the period-age relation using the 
Girardi et al. (1995) cluster age data. 

Table 1 gives the designations, UBV data, and ages of select clusters
in the LMC, and the names and periods of Cepheid variables that appear
to be associated with these clusters.  
The UBV data are from Bica et
al. (1996), and the ages come from the S values as determined in
Girardi et al.	(1995).  The sources of the data on the
cluster-Cepheid associations are indicated by the references.  These
are the most certain associations in the LMC.  Many other Cepheids are
reasonably close to clusters too (Efremov 1978, 1989), but their memberships
or ages were not judged to be as certain.  Sometimes it was difficult
to select unambiguous associations in the rich fields, as was the case
for the NGC 1850 binary cluster and a half dozen Cepheids close to it. 

We have not included the Cepheids with small amplitudes and sinusoidal
curves (subtype Cs), which are either the first-overtone pulsators or
those at the first crossing of the instability strip (Efremov 1968).
Their periods do not correspond to ages in the same way as for the
common Cepheids.

\begin{figure}
\vspace{2.3in}
\includegraphics{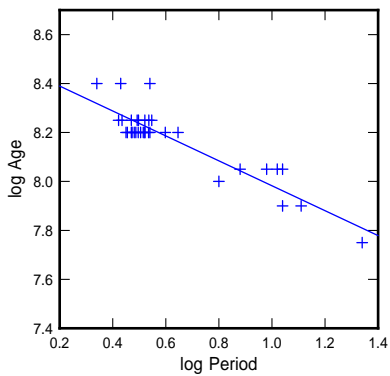}
\caption{Cepheid periods $P$ versus the ages $t$ of the clusters
in which the Cepheids are found, for Cepheids in the LMC.}
\label{fig:tp}
\end{figure}

Most of the Cepheids in the table are well inside cluster cores,
whereas those with references to Efremov (1978) are within 2 - 3 radii
of cores.  The latter is presumably the case for the Alcock et al.
(1995) data as well.

The data in table 1 gives the new period $P$ - age $t$ relation
shown in figure \ref{fig:tp}.
The correlation is
\begin{equation}
\log t = 8.492 - 0.509\log P,
\end{equation}
for $t$ in years and $P$ in days. The correlation coefficient
is 0.90. 
The result for paper 1 was $\log t = 8.157 - 0.677\log P$, which gives
shorter ages by a factor of $\sim3$.

\begin{figure}
\vspace{2.30in}
\includegraphics{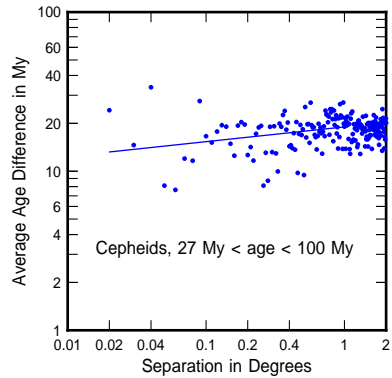}
\caption{Average age differences versus separation for Cepheids in the
LMC. There is no obvious correlation as there is for clusters, presumably 
because Cepheid stars
drift from their points of origin more rapidly
than clusters, and because the Cepheid ages are less accurate.}
\label{fig:ceph}
\end{figure}

The age differences between all pairs of Cepheids in the
Artyukhina et al. (1995) catalog with ages less than 100 Myr were
determined for intervals of deprojected
separation equal to 0.01 degree, and
the average age differences 
for each interval were found.  Figure \ref{fig:ceph} shows the 
average age difference versus separation for these Cepheids.
The figure 
suggests there is no significant 
correlation between $\Delta t$ and $S$ in the Cepheid data,
but the statistical uncertainty is large (out of 1200 Cepheids in the
Artyukhina et al. catalog, there are only 167 younger than 100 Myr, and among
these, there are only 1357 pairs with separations less than 1$^\circ$).

\begin{figure}
\vspace{1.5in}
\includegraphics{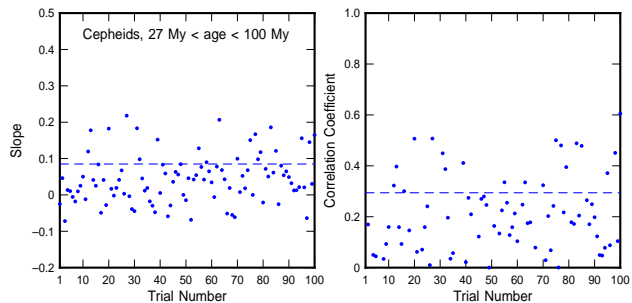}
\caption{The slopes (left) and correlation coefficients (right) for randomly
scrambled age data in the Cepheid sample are shown as points, while the
measured slope and correlation coefficient are shown as dashed lines.
The correlation coefficient for the real data is as poor as it is 
for a random
sample, and the slope of the $\Delta t-S$ regression for the real data
is within the
range of the noise for the random data.}
\label{fig:cephrandom} 
\end{figure}

Figure \ref{fig:cephrandom} shows 100 random trials for the Cepheids, 
using the observed Cepheid positions with randomly scrambled
ages, as for the clusters. The slopes of the $\Delta t-S$ relations
are shown on the left and the correlation coefficients are shown on 
the right. The dashed lines are the measured slope and correlation
coefficient, using the real Cepheid data. Evidently, the observed 
relation for Cepheids is consistent with noise. 
If there is a $\Delta t-S$ relation for these stars at birth, 
as appears to be the case for clusters, 
then this relation is apparently 
destroyed or obscured by the time the stars become Cepheids.

Possible reasons for the elimination of an initial
$\Delta t-S$ relation include random stellar motions
and inaccuracies in the Cepheid period-age relation.
The correlations shown here are better for clusters, perhaps because clusters 
have smaller space velocities than individual stars,
so they drift less from their points of origin than Cepheids.
Smaller space velocities for clusters might be reasonable if
clusters, born in cloud cores, start with the centroid
velocities of the star-forming clouds, while individual
stars, dispersed in associations, start with these centroid
velocities plus the dispersal speeds.  

\section{Discussion}

The positions and ages of clusters in the LMC suggest that there is a
correlation between the duration of star formation in a region and
the size of the region.  This correlation 
may be the result of star formation inside cloud complexes
that always live for several turbulent crossing times.

\begin{figure}
\vspace{3.6in}
\includegraphics{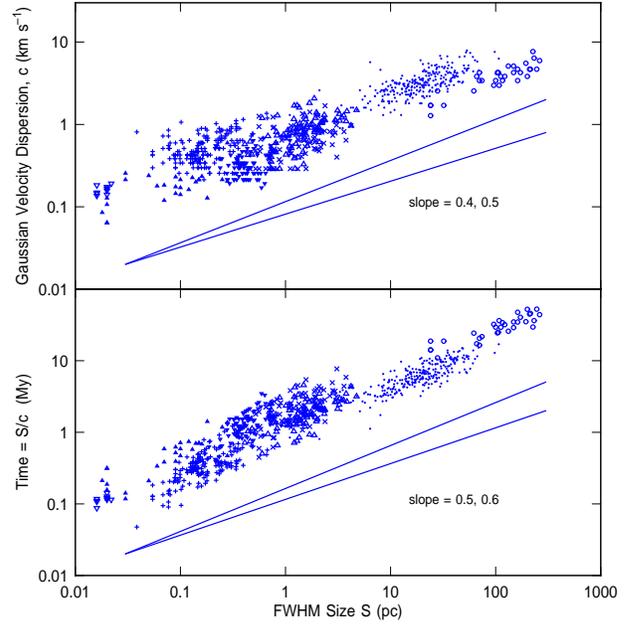}
\caption{Size and linewidth data from molecular cloud
surveys. Symbols are for giant molecular cloud
surveys = {\it dots}: Solomon et al. (1987),
{\it open circles}: Dame et al. (1986);
quiescent clouds = {\it filled triangles}: Falgarone et al. (1992),
{\it open
triangles}: Williams et al. (1994; the Maddalena-Thaddeus cloud),
{\it inverted open triangles}: Lemme et al. (1995; L1498),
{\it inverted filled triangles}: Loren (1989; Ophiuchus),
OB associations = {\it crosses}: Williams et al. (1994; Rosette),
{\it plus signs}: Stutzki \& G\"usten (1990; M17).}
\label{fig:sl}
\end{figure}

The turbulent crossing time 
in a region of star formation may be estimated from the size-linewidth relation
for the clouds that form stars. 
A compilation of the 
size-linewidth relations for
molecular clouds in the Milky Way is shown in
figure \ref{fig:sl}.
The size $S$ is the FWHM of clouds and clumps in various
surveys, and the linewidth $c$ is the Gaussian dispersion.
The average relation for all the surveys is
\begin{equation}
c(km\;s^{-1})\sim 0.7S(pc)^{0.5}.
\label{eq:sl}
\end{equation}
The ratio of $S$ to $c$ is shown on the bottom of the
figure.  Half of this ratio gives the turbulent crossing time,
\begin{equation} t_{crossing}(Myr)\sim{ {0.5S(pc)} \over {c(km\;s^{-1})}
} \approx 0.7 S(pc)^{0.5}.
\label{eq:mcs}\end{equation}
Because of the size-linewidth relation, the turbulent crossing time
increases with cloud size approximately as the square root. 
This slope is comparable to, although slightly larger than, the
$\Delta t-S$ relation for clusters, suggesting that the LMC cluster
relation results in part from turbulence. 
For separations of 10 pc and 100 pc, the average age difference between 
clusters in the LMC is 3.2, and 2.1 times the crossing time of 
galactic molecular clouds on the same scale. 
This suggests that star formation always proceeds with a time scale of 
$\sim2.5$ crossing times in the progenitor cloud.

The size-linewidth relation for clouds in the LMC is not well known,
but a recent study suggests that the linewidths in the L48 clumps 
are slightly larger for the clump size than the linewidths in 
galactic GMCs  (Kutner et al. 1998).  This could be a region of
unusually high pressure and not normal for star formation in the LMC, but
if this result is representative, then $t_{crossing}(S)$ would be 
slightly lower than for Galactic clouds, and the ratio of the
star formation time to the crossing time slightly larger than $\sim2.5$. 

The $\Delta t-S$ relation for clusters is not the result of OB association 
expansion or stellar drift 
at a fixed initial velocity because then the slope of the
correlation would be 1 instead of $\sim0.3$. It has to result from 
stellar or gas kinematics with velocity dispersions that scale with
a fractional power of the size of the region.

There are several important implications 
for star formation of the cluster $\Delta t-S$ relation:

1.  On all scales over which the correlation exists, small regions
come and go while large regions continue to form stars.  This implies
there might be some recycling of small-scale
star-forming material during the
lifetime of the larger region. Then 
metal enrichment from supernovae can be greater in the most active clouds,
and more variable from cloud to cloud than previously thought 
(Elmegreen 1998),
and total cloud ages can be longer than previously 
determined from cloud disruption times following only one epoch of
star formation (Leisawitz et al. 1989).

2.  Larger star-forming 
regions have both larger velocity dispersions and larger average
ages than smaller regions.  This trend is similar to that
found for OB subgroups and whole OB associations and may contribute to
the impression that OB subgroups expand into OB associations. 
There may not be this much expansion, however. Instead, there could 
be a
difference in the sizes and velocity dispersions of the two 
types of regions from birth.
OB subgroups are born small and they may stay moderately small
during the formation time of the
other subgroups.  All of the subgroups together define the
association, which is a composite of clumpy subparts. 

3.  The largest regions of star formation in giant spiral galaxies,
regions like Gould's Belt and other Star Complexes measuring 300 pc to
1 kpc, take so long to form (30 Myr) that they are not particularly
bright on average.  They are also not unified in appearance by a
single bright HII region or concentration of O stars, because the
general population of O stars that formed there has already aged off
the main sequence.  Instead, they are visible primarily as 
concentrations of Cepheid variables and other supergiant stars,
which is how
they were originally discovered (Efremov 1979, 1989).  Most of the O
stars are visible only in smaller concentrations, which appear 
as multiple cores inside the star complexes.
Thus the largest regions of star formation in giant spiral galaxies are often
overlooked, especially in H$\alpha$ or UV studies. 

This situation changes in smaller galaxies, where the star formation 
length and time scales are
generally shorter than in large galaxies (Elmegreen et al. 1996). 
In small galaxies, the largest regions of star formation can form
so quickly that there are still many OB stars, and then they appear
very bright, like 30 Dor (see review in Elmegreen \& Efremov 1998). 

4.  Regions of star formation that are defined by HII regions in
H$\alpha$ images of other galaxies tend to be concentrations of O-type
stars, and therefore have ages of around 10 million years.
These regions are the classical OB associations.  Because of the
$\Delta t-S$ correlation, they have a characteristic size that
corresponds to their age. For a measured characteristic size of 
OB associations equal to $\sim80$
pc (Lucke \& Hodge 1970; Efremov, Ivanov, \& Nikolov 1987),
equation \ref{eq:tspc} confirms that their duration of star formation
is $\sim14$ Myr.  
The identification of these regions is entirely based on the
presence of O-stars and bright emission nebulae, and is therefore only a
selection of one particular scale out of a continuum of scales for the
star-formation process.  This was implicitly the case in Efremov et
al.  (1987) and Battinelli et al.  (1996), where stars in M31 were
selected to be brightest in U or B to detect the O-associations.
However, OB associations are not representative of the star formation
process in general; they are only one level in a continuous hierarchy
of self-similar processes that extends from parsec to kiloparsec
scales (Elmegreen \& Efremov 1996, 1998).

\begin{figure}
\vspace{3.1in}
\includegraphics{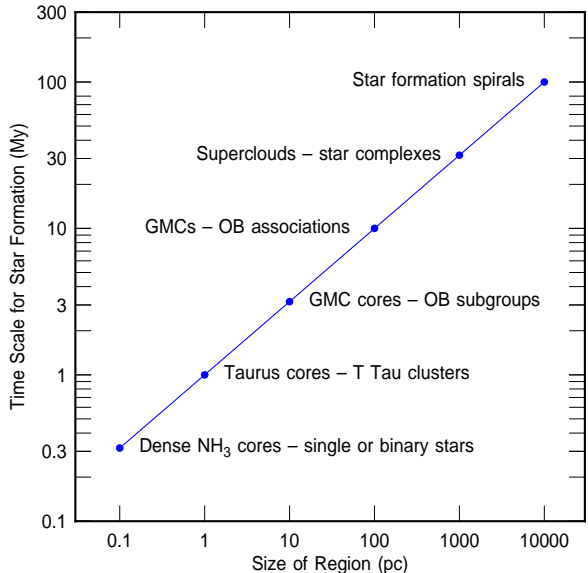}
\caption{Schematic diagram showing the size dependence of the 
duration of star formation in various regions.}
\label{fig:hier}
\end{figure}

Figure \ref{fig:hier} 
shows a schematic diagram of the sizes and durations of star
formation in regions that are commonly discussed.  The $\Delta t-S$
relation here is based on the square root approximation 
to the power in the size-linewidth relation (cf. Eq. \ref{eq:sl})
and not on the slightly shallower power-law dependence
for clusters written in equation (\ref{eq:tspc}).
The square root dependence gives a more sensible result over the whole
range of star-forming scales.
The figure considers 
small clumps in which individual stars form on
$10^5$ year time scales, T-Tauri stellar associations, 
OB subgroups and associations, and star
complexes, in which stars typically form on 1, 3, 10 and 30 Myr time scales, 
respectively. The figure also includes 
spiral arm segments (not density waves) that may extend for $\sim10$ kpc.  The
ages of spiral arm segments come from the pitch angles ($\sim
15^\circ$) and the rate of shear in the galaxy.  We
discussed previously how such star-forming spiral arms, which are
clearly distinct from density waves by their short lengths and lack of
dustlanes, are a natural extension of the star formation process to
scales larger than the disk thickness (Elmegreen \& Efremov 1996).

5. Star formation may be intimately connected with turbulence because
of the similarity between the $\Delta t-S$ relation and the
$S/c-S$ relation for molecular clouds. If true, then
turbulence would contribute to the rate of star formation in a region.
The formation of cloud structure by turbulent processes is well    
recognized (e.g., Langer et al. 1995; Falgarone \& Phillips 1996).

\section{Acknowledgements}

Comments by the referee on statistical effects in the Cepheid data
are greatly appreciated.
Yu.E. was partially supported by the RFBR grant 97-02-17358.

\begin{table*}
 \centering
 \begin{minipage}{140mm}
  \caption{Cepheids in LMC Clusters with Integral UBV Photometry}
  \begin{tabular}{@{}lrrrrrrrl@{}}  
cluster & V & U-B & B-V & S & $\log t$ & Cepheid 
& $\log_{10}P$(days) & ref.\\
NGC 1755 & 9.85 &-0.20 &0.16 &24 &8.00 & -  & 0.80 & Mateo 1992\\
NGC 1756 &12.24 & 0.09 &0.40 &30 &8.40 & -  & 0.34 & Alcock et al 1995\\
	 &	&      &     &	 &     & -  & 0.43 &\\
	 &	&      &     &	 &     & -  & 0.54 &\\
NGC 1866 & 9.73 &-0.02 &0.25 &28 &8.25 & HV 12197 & 0.497 & Welch et al. 1991\\
         &      &      &     &   &     &    12198 & 0.547 &\\
	 &	&      &     &	 &     &    12199 & 0.422 &\\
	 &	&      &     &	 &     &    12200 & 0.435 &\\
	 &	&      &     &	 &     &    12202 & 0.492 &\\
	 &	&      &     &	 &     &    12203 & 0.470 &\\
	 &	&      &     &	 &     &    12204 & 0.536 &\\
	 &	&      &     &	 &     &    V4	  & 0.521 &\\
NGC 2010 &11.72 &-0.07 &0.24 &27 &8.20 & HV  2599 & 0.455 & Gascoigne \& Hearnshaws 1971\\
	 &	&      &     &	 &     &    var3  & 0.54  &\\
NGC 2031 &10.83 &-0.07 &0.26 &27 &8.20 &	1 & 0.487 & Bertelli et al. 1993\\
	 &	&      &     &	 &     &	2 & 0.646 &\\
	 &	&      &     &	 &     &	3 & 0.598 &\\
	 &	&      &     &	 &     &	4 & 0.535 &\\
	 &	&      &     &	 &     &	5 & 0.521 &\\
	 &	&      &     &	 &     &	6 & 0.481 &\\
	 &	&      &     &	 &     &	7 & 0.496 &\\
	 &	&      &     &	 &     &	8 & 0.515 &\\
	 &	&      &     &	 &     &	9 & 0.470 &\\
	 &	&      &     &	 &     &       10 & 0.521 &\\
	 &	&      &     &	 &     &       11 & 0.450 &\\
	 &	&      &     &	 &     &       13 & 0.505 &\\
	 &	&      &     &	 &     &       14 & 0.473 &\\
NGC 2136 &10.54 &-0.13 &0.28 &25 &8.05 & HV  2868 & 0.88  & Robertson 1974, Efremov 1978\\
	 &	&      &     &	 &     &     2870 & 0.98  &\\
	 &	&      &     &	 &     &    12230 & 1.02  &\\
	 &	&      &     &	 &     &      B21 & 1.04  &\\
NGC 2214 &10.93 &-0.27 &0.11 &23 &7.90 &       B1 & 1.04  & Robertson 1974\\
SL 106	 &11.28 &-0.33 &0.15 &21 &7.75 &   HV2245 & 1.34  & Efremov 1978\\
SL 234	 &12.44 &-0.36 &-0.03&23 &7.90 &   HV2321 & 1.11  & Efremov 1978
\end{tabular}
\end{minipage}
\end{table*}

\bsp

\label{lastpage}

\end{document}